\documentclass[a4paper,11pt]{article}
\pdfoutput=1 % if your are submitting a pdflatex (i.e. if you have images in pdf, png or jpg format)
\usepackage{jcappub} 
\usepackage[T1]{fontenc} 
\usepackage{amssymb,amsmath,amsfonts,mathrsfs}
\usepackage{hyperref}  
\usepackage{float}
\usepackage{xcolor}
\usepackage{tensor}
\usepackage{tikz}
\usetikzlibrary{arrows,decorations.pathmorphing,decorations.markings,patterns,decorations.pathreplacing}
\tikzset{
	graviton/.style={decorate,
		decoration={coil,amplitude=2pt, segment length=2pt}} 
}

\title{Non-perturbative quantum Galileon in the exact renormalization group}
\author{Christian F. Steinwachs}
\affiliation{Physikalisches Institut, Albert-Ludwigs-Universit\"at Freiburg, Hermann-Herder-Str.~3, 79104 Freiburg, Germany}
\emailAdd{christian.steinwachs@physik.uni-freiburg.de}
\abstract{
We investigate the non-perturbative renormalization group flow of the scalar Galileon model in flat space. We discuss different expansion schemes of the Galileon truncation, including a heat-kernel based derivative expansion, a vertex expansion in momentum space and a curvature expansion in terms of a covariant geometric formulation. We find that the Galileon symmetry prevents a quantum induced renormalization group running of the Galileon couplings. Consequently, the Galileon truncation only features a trivial Gaussian fixed point.}
\begin{document}
\allowdisplaybreaks[1]
\maketitle
\flushbottom
\section{Introduction}

The Galileon in flat space is a higher-derivative theory of a single scalar field which arises from the decoupling limit of the Dvali-Gabadadze-Porrati model \cite{Dvali2000} and whose particular symmetry structure ensures the second-order character of the field equations and the absence of ghost-like excitations \cite{Nicolis2004,Nicolis2009}. Its generalization to the covariant Galileon in curved space, Horndeski's scalar-tensor theory \cite{Horndeski:1974wa,Deffayet:2009wt}, has many applications in cosmological models of the early and late time acceleration of the universe, see e.g.~\cite{Chow:2009fm,Gannouji:2010au,Creminelli:2010ba,DeFelice:2010pv,Kobayashi:2010cm,Burrage:2010cu,Deffayet:2011gz,Gubitosi:2012hu,Renaux-Petel:2013ppa}.

Various aspects of the classical and quantum properties of higher-derivative scalar effective field theories including the Galileon, the Dirac-Born-Infeld model and the non-linear sigma model have been studied \cite{deRham:2010eu,DeRham2014,Pirtskhalava2015}, including the perturbative renormalization of the Galileon  \cite{Nicolis2004,PaulaNetto2012,Rham2013,Heisenberg2014a,Brouzakis2014,Brouzakis2014a,Goon:2016ihr,Saltas2017,Heisenberg2019, Heisenberg:2019wjv,Goon:2020myi} and of the non-linear sigma model \cite{Howe:1986vm,Buchbinder:1991jw,Buchbinder:1988ei,Barvinsky2018}, as well as on-shell scattering amplitudes in effective scalar field theories \cite{Kampf2013,Kampf2014,Cheung2015,Cheung2015a,Cheung2016,Cheung2017,Carrasco:2019qwr, Kampf:2019mcd,Kampf:2020tne}. 

In this article we study the non-perturbative renormalization group (RG) flow of the Galileon truncation in the framework of the exact renormalization group equation (ERGE); for a recent review on the ERGE see \cite{Dupuis:2020fhh}.
The scalar nature of the Galileon field also permits to focus on the structural aspects of higher-derivative theories in the ERGE. In particular, it permits to investigate the dependence of the RG properties on the regulator choice and to test various optimization criteria without having to deal with the additional complications and off-shell ambiguities present in gauge theories, see e.g.~\cite{Litim:2000ci,Litim:2001up,Codello:2013bra,Knorr:2020rpm,PhysRevLett.123.240604}.

Truncations of scalar field theories based on derivative expansions up to order $\mathcal{O}(\partial^6)$ have been studied previously in the in the context of the ERGE, see e.g.~\cite{Tetradis:1992qt,Morris:1994ie,Morris:1994ki,Morris:1997xj,Litim:2001dt,PhysRevB.68.064421,Percacci:2009dt,Percacci:2009fh,Codello:2018nbe,PhysRevD.98.016013,PhysRevLett.123.240604,Codello:2020lta}.
In \cite{Codello:2012dx}, the RG flow of $d$-dimensional surfaces embedded in a $d+1$-dimensional space and the correspondence to the Galileon, which emerges from this geometric formulation in the non-relativistic limit, has been studied.

In contrast, in the present work, we directly investigate the non-perturbative RG flow of the Galileon truncation in different expansion schemes, including a derivative expansion up to $\mathcal{O}(\partial^8)$, a vertex expansion in momentum space up to $\mathcal{O}(\pi^5)$, and a curvature expansion within a compact covariant geometric formulation in terms of an effective Galileon metric which results from a resummation of infinitely many operators with a fixed number of derivatives per field.
We find that the non-renormalization property of the Galileon in the perturbative quantization \cite{Luty:2003vm,Hinterbichler:2010xn,Goon:2016ihr,Heisenberg2019a, Heisenberg:2019wjv} carries over to the non-perturbative RG flow. That is, the Galileon beta functions do not receive any quantum-induced contributions and are only driven by the classical (canonical) scaling dimensions of the Galileon couplings, which implies that the RG system only features a trivial Gaussian fixed point. Nevertheless, compared to the perturbative calculation of the one-loop divergences \cite{PaulaNetto2012,Brouzakis2014,Brouzakis2014a,Heisenberg2019,Heisenberg:2019wjv}, a larger set of operators, which lead out of the Galileon truncation, is induced in the functional trace of the Wetterich equation. These terms correspond to power-law divergences and are absent in dimensional regularized perturbative calculations.

The article is structured as follows:
In Sect.~\ref{ERG} we introduce the required formalism for the non-perturbative quantization of the Galileon within the Exact Renormalization Group Equation (ERGE).
In Sect.~\ref{ExpSchemes} we calculate the functional trace in the Wetterich equation by three different expansion schemes, which are based on a $\mathcal{O}(\partial^8)$ derivative expansion, a vertex expansion in momentum space and a covariant geometric resummation techniques. We discuss the implications of our calculation in Sect.~\ref{Discussion} and conclude in Sect.~\ref{Conclusion}.
\section{The Galileon truncation in the exact functional renormalization group}
\label{ERG}

The exact renormalization group (RG) is defined by the Wetterich equation \cite{Wetterich1993,Reuter1994,Morris1994}
\begin{align}
\dot{\Gamma}_k=\mathrm{Tr}\left(\frac{\dot{\mathfrak{R}}_k}{\Gamma_k^{(2)}+\mathfrak{R}_k}\right).\label{Wetteq}
\end{align}
The Euclidean effective averaged action (EAA) $\Gamma_k[\pi]$ is a functional of the Galileon field $\pi$ and depends on the abstract RG scale $k$. It approaches the bare action  $\Gamma_{\infty}=S$ in the limit $k\to\infty$ and the effective action $\Gamma_{0}:=\Gamma$ in the limit $k\to0$. A dot denotes differentiation with respect to the logarithmic $k$-derivative $\dot{X}_k:=\partial_{t}X_k:=k\partial_kX_k$. The presence of the regulator $\mathfrak{R}_k$ in \eqref{Wetteq} provides an infrared (IR) cutoff, while its scale derivative $\dot{\mathfrak{R}}_k$ provides an ultraviolet (UV) cutoff such that the result of the trace is both IR and UV finite.\footnote{The regulator enters the action in the path integral quadratic in the fields and hence acts as an effective $k$-dependent mass term $\mathfrak{R}_k\propto k^2$  for fluctuations with $p^2\ll k^2$ while it satisfies $\mathfrak{R}_k\approx 0$ for $p^2\gg k^2$ ($\mathfrak{R}_k\to 0$ in the limit $k\to0$).}
The one-loop structure of \eqref{Wetteq} is reflected by the functional trace. The  Galileon truncation of the EAA in $d=4$ dimensional flat space with Euclidean metric $\delta_{\mu\nu}=\text{diag}(1,1,1,1)$ reads
\begin{align}
\Gamma_k[\pi]={}&\sum_{i=2}^{5}\Gamma^{(i)}_k[\pi],\label{Gact}\\
\Gamma_k^{(2)}[\pi]={}&-\frac{c_2}{12}\int\mathrm{d}^{4}x\,\pi\epsilon^{\mu\nu\rho\sigma}\tensor{\epsilon}{^{\alpha}_{\nu\rho\sigma}}\pi_{\mu\alpha},\label{Gam1}\\
\Gamma_k^{(3)}[\pi]={}&\frac{c_3}{12}\int\mathrm{d}^{4}x\,\pi\epsilon^{\mu\nu\rho\sigma}\tensor{\epsilon}{^{\alpha\beta}_{\rho\sigma}}\pi_{\mu\alpha}\pi_{\nu\beta},\\
\Gamma_k^{(4)}[\pi]={}&-\frac{c_4}{8}\int\mathrm{d}^{4}x\,\pi\epsilon^{\mu\nu\rho\sigma}\tensor{\epsilon}{^{\alpha\beta\gamma}_{\sigma}}\pi_{\mu\alpha}\pi_{\nu\beta}\pi_{\rho\gamma},\\
\Gamma_k^{(5)}[\pi]={}&\frac{c_5}{10}\int\mathrm{d}^{4}x\,\pi\epsilon^{\mu\nu\rho\sigma}\tensor{\epsilon}{^{\alpha\beta\gamma\delta}}\pi_{\mu\alpha}\pi_{\nu\beta}\pi_{\rho\gamma}\pi_{\sigma\delta}\label{Gam4}\,.
\end{align}
The symmetric tensor $\pi_{\mu\nu}$ is defined as $\pi_{\mu\nu}:=\partial_{\mu}\partial_{\nu}\pi$ and $\varepsilon^{\mu\nu\rho\sigma}$ is the totally antisymmetric Levi-Civita symbol.
The Galileon field and the partial derivatives have mass dimension $[\pi]=[\partial_{\mu}]=1$, while the running couplings $c_i$, $i=2,\ldots 5$ have mass dimension\footnote{We suppress the subscript $k$ that indicates the RG scale dependence.}
\begin{align}
[c_i]=-3(i-2).\label{MassDim}
\end{align}
Since the Galileon action only involves derivative interactions, it is obviously invariant under shift symmetries $\pi\to\pi+\lambda$ with a constant $\lambda$. However, the action \eqref{Gact} is even invariant under the more restrictive Galileon transformations with a constant vector $v_{\mu}$,
\begin{align}
\pi\to\pi+\lambda+ v_{\mu}x^{\mu}.\label{Gsym}
\end{align}
The invariance \eqref{Gsym} leads to the particular structure of the Galileon interactions \eqref{Gam1}-\eqref{Gam4} and ensures that, despite the presence of higher derivative terms, the field equations are of second order and no ghost-like excitations appear in the spectrum.

Expanding the tensor contractions in \eqref{Gam1}-\eqref{Gam4} and integrating by parts, the Galileon truncation \eqref{Gact} is expressed in terms of a basis of invariants which does not contain self-contractions of derivatives acting on the same field (we use $\partial_{\mu_1\ldots\mu_n}=\partial_{\mu_1}\ldots\partial_{\mu_n}$),
\begin{align}
\Gamma_k=\frac{1}{2}\int\mathrm{d}^{4}x\,&\left\{c_2(\partial\pi)^2+c_3(\partial_{\mu}\pi)(\partial_{\nu}\pi)(\partial^{\mu\nu}\pi)+c_4\left[(\partial_{\mu}\pi)(\partial_{\nu}\pi)(\partial_{\rho}\pi)(\partial^{\mu\nu\rho}\pi)\right.\right.\nonumber\\
&\left.\left.+3(\partial_{\mu}\pi)(\partial_{\nu}\pi)(\partial^{\rho\mu}\pi)(\tensor{\partial}{_{\rho}^{\nu}}\pi)\right]+c_5\left[(\partial_{\mu}\pi)(\partial_{\nu}\pi)(\partial_{\rho}\pi)(\partial_{\sigma}\pi)(\partial^{\mu\nu\rho\sigma}\pi)\right.\right.\nonumber\\
&\left.\left.+12(\partial_{\mu}\pi)(\partial_{\nu}\pi)(\partial_{\rho}\pi)(\tensor{\partial}{^{\mu}_{\sigma}}\pi)(\tensor{\partial}{^{\sigma\nu\rho}}\pi)+12(\partial_{\mu}\pi)(\partial_{\nu}\pi)(\tensor{\partial}{^{\mu}_{\rho}}\pi)(\tensor{\partial}{^{\rho}_{\sigma}}\pi)(\tensor{\partial}{^{\sigma\nu}}\pi)\right]\right\}.\label{AEAGalExp}
\end{align}
We perform the linear split of the Galileon field into background plus perturbation
\begin{align}
\pi(x)=\bar{\pi}(x)+\delta\pi(x).
\end{align}
We identify the physical mean field $\langle\pi\rangle$ with $\bar{\pi}$ and omit the bar indicating a background quantity. 
The scalar second-order fluctuation operator $F(\partial)$ is defined by the Hessian 
\begin{align}
\left.\frac{\delta^2\Gamma_{k}[\pi]}{\delta\pi(x)\delta\pi(x')}\right|_{\pi}=F(\partial)\delta(x,x')= -\left(\mathcal{G}^{-1}\right)^{\mu\nu}\partial_{\mu}\partial_{\nu}\delta(x,x').\label{Op}
\end{align}
The structure of the fluctuation operator in \eqref{Op} suggests to identify $\left(\mathcal{G}^{-1}\right)^{\mu\nu}$ with the inverse of an ``effective Galileon metric'' and is explicitly defined in terms of $\pi$
\begin{align}
\left(\mathcal{G}^{-1}\right)^{\mu\nu}:={}&\left(\mathcal{G}^{-1}_{(0)}\right)^{\mu\nu}+\left(\mathcal{G}^{-1}_{(2)}\right)^{\mu\nu}+\left(\mathcal{G}^{-1}_{(4)}\right)^{\mu\nu}+\left(\mathcal{G}^{-1}_{(6)}\right)^{\mu\nu},\label{GInvExpl}\\
\left(\mathcal{G}^{-1}_{(0)}\right)^{\mu\nu}={}&\frac{c_2}{6}\varepsilon^{\mu\alpha\rho\sigma}\varepsilon^{\nu}_{\alpha\rho\sigma}=c_2g^{\mu\nu},\\
\left(\mathcal{G}^{-1}_{(2)}\right)^{\mu\nu}={}&-\frac{c_3}{2}\varepsilon^{\mu\alpha\rho\sigma}
\tensor{\varepsilon}{^{\nu\beta}_{\rho\sigma}}\pi_{\alpha\beta}=c_3\left[\partial^{\mu}\partial^{\nu}\pi+g^{\mu\nu}\Delta\pi\right],\label{Ginv2}\\
\left(\mathcal{G}^{-1}_{(4)}\right)^{\mu\nu}={}&\frac{3}{2}c_4\varepsilon^{\mu\alpha\rho\sigma}
\tensor{\varepsilon}{^{\nu\beta\gamma}_{\sigma}}\pi_{\alpha\beta}\pi_{\rho\gamma}=\frac{3}{2}c_4\left[2(\partial^{\mu}\partial_{\rho}\pi)(\partial^{\rho\nu}\pi)+2(\partial^{\mu\nu}\pi)(\Delta\pi)\right.\nonumber\\
&\left.+g^{\mu\nu}(\Delta\pi)^2-g^{\mu\nu}(\partial_{\rho}\partial_{\sigma}\pi)(\partial^{\rho}\partial^{\sigma}\pi)\right],\label{Ginv4}\\
\left(\mathcal{G}^{-1}_{(6)}\right)^{\mu\nu}={}&-2c_5\varepsilon^{\mu\alpha\rho\sigma}\tensor{\varepsilon}{^{\nu\beta\gamma\kappa}}\pi_{\alpha\beta}\pi_{\rho\gamma}\pi_{\sigma\kappa}=2c_5\left[3(\partial^{\mu}\partial^{\nu}\pi)(\Delta\pi)^2+6(\partial^{\mu}\partial_{\rho}\pi)(\partial^{\rho}\partial^{\nu}\pi)(\Delta\pi)\right.\nonumber\\
&\left.+6(\partial^{\mu}\partial_{\rho}\pi)(\partial^{\rho}\partial_{\sigma}\pi)(\partial^{\sigma}\partial^{\nu}\pi)-3(\partial^{\mu}\partial^{\nu}\pi)(\partial_{\rho}\partial_{\sigma}\pi)(\partial^{\rho}\partial^{\sigma}\pi)+g^{\mu\nu}(\Delta\pi)^3\right.\nonumber\\
&\left.-2g^{\mu\nu}(\partial_{\rho}\partial^{\sigma}\pi)(\partial_{\sigma}\partial^{\lambda}\pi)(\partial_{\lambda}\partial^{\rho}\pi)-3g^{\mu\nu}(\partial_{\rho}\partial_{\sigma})(\partial^{\rho}\partial^{\sigma})(\Delta\pi)\right].\label{Ginv6}
\end{align}
The subscript of the tensor $\left(\mathcal{G}^{-1}_{(i)}\right)^{\mu\nu}$ indicates the number of background derivatives and $\Delta:=-\delta^{\mu\nu}\partial_{\mu\nu}$ defines
 the positive definite Laplacian. The number of background derivatives defines the background derivative order (BDO) of a tensor. For example, we write (suppressing tensor indices) $\mathcal{G}^{-1}_{(i)}=\mathcal{O}\left(\partial^i\right)$ to indicate that the $\mathcal{G}^{-1}_{(i)}$ have BDO $i$.
The effective Galileon metric $\mathcal{G}_{\mu\nu}$ is defined as the inverse of \eqref{Op} via the relation
\begin{align}
\mathcal{G}_{\mu\rho}\left(\mathcal{G}^{-1}\right)^{\rho\nu}=\delta^{\nu}_{\mu}.\label{Gmetric}
\end{align}
The Galileon metric $\mathcal{G}_{\mu\nu}$ is assumed to be positive definite to ensure its non-degeneracy\footnote{Positive definiteness requires that all eigenvalues of $\mathcal{G}_{\mu\nu}$ must be positive. As the eigenvalues are in general functions of the running couplings positive definiteness implies constraints on the $c_i$.}
\begin{align}
\det(\mathcal{G})\neq 0.
\end{align}
The metric $\mathcal{G}_{\mu\nu}$ is not constant, but importantly its inverse $\left(\mathcal{G}^{-1}\right)^{\mu\nu}$ is divergence-free 
\begin{align}
\partial_{\mu}\left(\mathcal{G}^{-1}\right)^{\mu\nu}=0.\label{GinvDfree}
\end{align}
Due to the particular structure of $\left(\mathcal{G}^{-1}\right)^{\mu\nu}$ this also holds separately for the individual tensors $\left(\mathcal{G}^{-1}_{(i)}\right)^{\mu\nu}$, $i=0,2,4,6$. The regulated propagator is defined in terms of $\Gamma^{(2)}_k$ and $\mathfrak{R}_k$ by\footnote{We suppress the $k$-label indicating the dependence eon the RG scale.}
\begin{align}
G_{\mathfrak{R}}:=\frac{1}{F+\mathfrak{R}}.\label{RegProp}
\end{align}
In the next section, we calculate the functional trace in the Wetterich equation in different expansion schemes.
\section{Evaluation of the functional trace in different expansion schemes}
\label{ExpSchemes}
The Wetterich equation only provides an ``exact'' equation if the RG flow ``closes'', that is, if the operators that are arise in the functional trace of the flow equation \eqref{Wetteq} are of the same form as those present in the ansatz for the EAA. However, in general, the fluctuation operator derived from the Hessian of the EAA induces operators in the functional trace which lead out of the truncation. Consistency of the Wetterich equation then requires to neglect these operators in order for the flow to close at least approximately within the projection onto the finite truncation.

Therefore an important practical question is connected to the principles according to which the operators of a finite truncation should be chosen. In most cases, standard expansion schemes common to those in effective field theories (EFTs) provide a natural guiding principle. Any truncation is ultimately based on counting the number of derivatives, the number of fields, the powers of mass parameters, or any other small parameters such as $1/N$ in the large $N$ limit, or $\epsilon$ in the expansion around two-dimensional space $d=2+ \epsilon$, or a particular combination of these. The most agnostic power counting scheme may be viewed to correspond to truncations ordered by operators with increasing canonical mass dimension. However, in general, the choice of the best adapted expansion schemes depends on the underlying physical problem. In particular, in case the model under consideration has additional symmetries, more sophisticated combined expansion scheme may be more appropriate as e.g.~in the context of chiral perturbation theory \cite{Honerkamp:1971sh,Gasser:1983yg}.
\subsection{Derivative expansion}
We denote the functional trace in the Wetterich equation \eqref{Wetteq} for the ansatz \eqref{Gact} by
\begin{align}
T:=\mathrm{Tr}\left(\frac{\dot{\mathfrak{R}}}{F+\mathfrak{R}}\right).\label{TTrace}
\end{align}
We also define the regulated fluctuation operator $F_{\mathfrak{R}}=F+\mathfrak{R}$ and sort the individual terms in $F_{\mathfrak{R}}$ according to their BDO
\begin{align}
F_{\mathfrak{R}}={}&P+\Pi,\qquad P(\Delta)=c_2\Delta+\mathfrak{R}(\Delta),\qquad \Pi(\partial)=\left[\mathcal{G}^{-1}_{(2)}+\mathcal{G}^{-1}_{(4)}+\mathcal{G}^{-1}_{(6)}\right]^{\mu\nu}\partial_{\mu\nu}.\label{PiOp}
\end{align}
Following the classification scheme of \cite{Codello2009}, we choose a type I regulator $\mathfrak{R}$ without specifying any concrete profile function. 
We denote the BDO of the individual operators by
\begin{align}
F{}&=\mathcal{O}(\partial^0),& P&=\mathcal{O}(\partial^0),& \Pi&=\mathcal{O}(\partial^2),\nonumber\\
\mathcal{G}^{-1}_{(2)}{}&=\mathcal{O}(\partial^2), & \mathcal{G}^{-1}_{(4)}&=\mathcal{O}(\partial^4),& \mathcal{G}^{-1}_{(6)}&=\mathcal{O}(\partial^6).\label{BDOs}
\end{align}
In view of the BDOs in \eqref{BDOs}, we can perform a systematic derivative expansion in powers of the perturbation $\Pi$ by expanding the Greens function $G_{\mathfrak{R}}$ of $F_{\mathfrak{R}}$ up to the required order
\begin{align}
G_{\mathfrak{R}}=\frac{1}{F_{\mathfrak{R}}}=\frac{1}{P(\Delta)}-\frac{1}{P(\Delta)}\Pi(\partial)\frac{1}{P(\Delta)}+\frac{1}{P(\Delta)}\Pi(\partial)\frac{1}{P(\Delta)}\Pi(\partial)\frac{1}{P(\Delta)}+\ldots\label{Gexp}
\end{align}
For a $\mathcal{O}(\partial^8)$ truncation, we need to expand \eqref{Gexp} up to terms involving four powers of $\Pi$.
In order to explicitly evaluate the functional trace we need to arrange the resulting operators into a standard form by commuting all functions of the Laplacian to the very right. This can be accomplished iteratively by repeated use of the basic identity
\begin{align}
\left[\Pi,\frac{1}{P}\right]=\frac{1}{P}\left[P,\Pi\right]\frac{1}{P}.\label{Bcom}
\end{align}
In view of $\Pi=\mathcal{O}\left(\partial\right)$ and the fact that each commutator $[P,\Pi]$ increases the BDO at least by one, this expansion is guaranteed to be efficient.
We need to evaluate commutators $\left[P,\Pi\right]$ which involve an arbitrary function $P(\Delta)$ of  $\Delta$. These commutators are again expanded up to the required BDO order $r$ by using the general formula
\begin{align}
\left[P,\Pi\right]=\sum_{n=1}^{r}\frac{1}{n!}(-1)^{n}[\Pi,\Delta]_n\,P^{(n)}(\Delta).\label{CommLapPi}
\end{align}
Here,  $P^{(n)}(\Delta)$ is the $n$th derivative of $P(\Delta)$ with respect to the argument and all dependence on $\Delta$ has been ordered to the very right in \eqref{CommLapPi}. The $n$-fold commutator is defined iteratively 
\begin{align}
[\Pi,\Delta]_n:=[[\Pi,\Delta]_{n-1},\Delta], \qquad [\Pi,\Delta]_1:=[\Pi,\Delta],\qquad[\Pi,\Delta]_0:=\Pi.
\end{align}
Once the required BDO has been reached all factors of $1/P$ can freely be commuted to the right. The result of this algorithm is that the functional trace \eqref{TTrace}, truncated at $\mathcal{O}(\partial^8)$, reduces to a sum of functional traces of the form
\begin{align}
\mathrm{Tr}\left[B^{\mu_1\ldots\mu_j}\partial_{\mu_1}\ldots\partial_{\mu_j}W_{(i,j,k,l)}^{(m)}(\Delta)\right],\label{Btrace}
\end{align}
with  background-tensorial coefficients $B^{\mu_1\ldots\mu_j}$ and $\Delta$-dependent functions $W_{(i,j,k,\ell)}^{(m)}(\Delta)$,
\begin{align}
W_{(i,j,k,\ell)}^{(m)}=\left(P^{(1)}\right)^i\,\left(P^{(2)}\right)^j\,\left(P^{(3)}\right)^k\,\left(P^{(4)}\right)^\ell\,P^{-m}\,\dot{\mathfrak{R}}
\end{align}
In order to evaluate these functional traces explicitly, we perform the Mellin transform
\begin{align}
W(\Delta)=\int\mathrm{d}s\,\mathcal{L}^{-1}\left[W\right](s)\,e^{-s\Delta}.\label{Laplace}
\end{align}
In this way, the evaluation of the functional traces in \eqref{Btrace} reduce to the evaluation of the Universal Functional Traces (UFTs) \cite{Barvinsky1985} of the bare Laplacian $\Delta$,
\begin{align}
\mathscr{U}_{\mu_1\ldots\mu_j}(\Delta|s):={}&\left.\partial_{\mu_1}\ldots\partial_{\mu_j}e^{-s\Delta}\delta(x,y)\right|_{y=x}=\frac{1}{(4\pi s)^{2}}\,\partial_{\mu_1}\ldots\partial_{\mu_j}\left.e^{-\frac{(x-y)^2}{4s}}\right|_{y=x}.\label{UFTFlat}
\end{align}
In the last equality of \eqref{UFTFlat} we used the exact result for the heat-kernel of the Laplacian in $d=4$ dimensional flat space. The chain of derivatives in \eqref{UFTFlat} only leads to a combinatorial factor times the totally symmetrized product of metric tensors normalized such that each term carries unit weight
\begin{align}
[\delta_{\mathrm{sym}}]_{\mu_1\ldots\mu_{2j}}:=(2j-1)!!\,\delta_{(\mu_1\mu_2}\ldots\delta_{\mu_{2j-1}\mu_{2j})}\label{dsym}.
\end{align}
Due to the structure of \eqref{UFTFlat} only UFTs with an even number of indices $2j$ are non-zero,
\begin{align}
\mathscr{U}_{\mu_1\ldots\mu_{2j}}(\Delta|s)=\frac{(-1)^{j}}{(2\pi)^{2}}(2s)^{-(j+2)}[\delta_{\mathrm{sym}}]_{\mu_1\ldots\mu_{2j}}.\label{flatUFTClosed}
\end{align}
We absorb the $s$-integral into the Q-functionals defined for an arbitrary function $W(\Delta)$,\footnote{Note that the definition of the $Q$-functional is slightly different from the standard definition as it involves an additional numerical factor of $2^{-n}/4\pi^2$.}
\begin{align}
Q_{n}[W]:={}&\frac{2^{-n}}{(2\pi)^{2}}\int_0^{\infty}\mathrm{d}s\,s^{-n}\mathcal{L}^{-1}\left[W\right](s). \label{Qintflat}
\end{align}
Here, $\mathcal{L}^{-1}[W](s)$ is the inverse Laplace transform of $W[z]$ and the $Q$-functional can be expressed in terms of the original function $W[z]$ by
\begin{align}
Q_{n}[W]=\frac{2^{-n}}{(2\pi)^{2}}\frac{1}{\Gamma(n)}\int_0^{\infty}\mathrm{d}z\, z^{n-1}W(z).\label{QOrig}
\end{align}
The algorithm \eqref{TTrace}-\eqref{QOrig} is similar to that proposed in \cite{Benedetti2011} but differs in the expansion which is based on a derivative expansion and not on an expansion of curvatures or background mass dimension. 
 
Sometimes, it might happen that different expansion schemes accidentally fall together such as in the context of gravity. In this case diffeomorphism invariance suggests a manifestly covariant expansion which runs in powers of (derivatives of) curvature invariants. In view of the non-linear relation between the curvature and the metric (and its derivatives), a vertex expansion in powers of the metric perturbations breaks this manifest covariance. As the metric field is dimensionless (for coordinates having the physical dimension of length), a curvature expansion might also be viewed as a derivative expansion $R\sim \partial^2g$. However, in view of the diffeomorphism symmetry, a more appropriate viewpoint is to consider the expansion in curvatures as a covariant vertex expansion. A true covariant derivative expansion would then count derivatives acting on curvatures and not on metric perturbations. In certain situations, even a resummation of terms is possible and adequate. In the aforementioned context of the covariant perturbation theory, this corresponds to resum all operators with a fixed number of curvatures but an arbitrary number of derivatives leading to non-local form factors \cite{Barvinsky1990,Barvinsky:1990uq,Codello:2012kq}. Recently, such truncations have also been studied in the context of the ERGE \cite{Codello:2015oqa,Knorr:2019atm}. Similarly, the Galileon in flat spacetime can be reformulated in geometrical terms \cite{Heisenberg2019a} and thereby permits a resummation of an infinite number of operators with at least two powers of derivatives per field into curvature tensors, which are defined with respect to an effective Galileon metric as we will discuss in more detail in Sect.~\ref{CovGeoSum}.

Combining the expansions \eqref{Gexp}, \eqref{Bcom} and \eqref{CommLapPi} with \eqref{Laplace}, \eqref{flatUFTClosed} and \eqref{Qintflat}, the $\mathcal{O}(\partial^8)$ derivative expansion of the functional trace \eqref{TTrace} acquires the form
\begin{align}
T=\frac{1}{2}\int\mathrm{d}^4x{}&\left\{C_0
+C_{24}\,(\partial_{\mu\nu}\pi)(\partial^{\mu\nu}\pi)+C_{26} (\partial_{\mu\nu\rho}\pi)(\partial^{\mu\nu\rho}\pi)+ C_{36a}\, (\partial_{\mu\nu\rho}\pi)(\partial^{\mu\nu}\pi)(\partial^{\rho}\pi)\right.\nonumber\\
&+ C_{36b}\, (\partial_{\mu}\partial^{\nu}\pi)(\partial_{\nu}\partial^{\rho}\pi)(\partial_{\rho}\partial^{\mu}\pi)+C_{28} (\partial_{\mu\nu\rho\sigma}\pi)(\partial^{\mu\nu\rho\sigma}\pi)\nonumber\\
&
+C_{38a}\,(\partial_{\mu\nu\rho}\pi)(\partial^{\rho}\partial_{\sigma}\pi)(\partial^{\mu\nu\sigma}\pi)
+ C_{38b}\,(\partial_{\mu\nu}\pi)(\partial^{\mu\nu\rho\sigma}\pi)(\partial_{\rho\sigma}\pi)\nonumber\\
&+C_{38c}\, (\partial_{\mu\nu\rho\sigma}\pi)(\partial^{\mu\nu\rho}\pi)(\partial^{\sigma}\pi)+C_{48a}\, (\partial_{\mu\nu}\pi)(\partial^{\mu\nu}\pi)(\partial_{\rho\sigma}\pi)(\partial^{\rho\sigma}\pi)\nonumber\\
&+ C_{48b}\,(\partial_{\mu}\partial^{\nu}\pi)(\partial_{\nu}\partial^{\rho}\pi)(\partial_{\rho}\partial^{\sigma}\pi)(\partial_{\sigma}\partial^{\mu}\pi)+C_{48c}\,(\partial_{\mu\nu\rho\sigma}\pi)(\partial^{\mu\nu}\pi)(\partial^{\rho}\pi)(\partial^{\sigma}\pi)\nonumber\\
&+C_{48d}\,(\partial_{\mu\nu\rho}\pi)(\partial^{\mu\nu\sigma}\pi)(\partial^{\rho}\pi)(\partial_{\sigma}\pi)+C_{48e}\,(\partial_{\mu\nu\rho}\pi)(\partial^{\mu\sigma}\pi)(\partial_{\sigma}\partial^{\nu}\pi)(\partial^{\rho}\pi)\nonumber\\
&\left.+C_{48f}\,(\partial_{\mu\nu\rho}\pi)(\partial^{\mu\nu}\pi)(\partial^{\rho\sigma}\pi)(\partial_{\sigma}\pi) \right\}+\mathrm{t.d.}+\mathcal{O}\left(\partial^9\right).\label{FinalTraceT8}
\end{align}
In the final result \eqref{FinalTraceT8}, we have used the explicit expressions \eqref{Ginv2}-\eqref{Ginv6} for the $\mathcal{G}^{-1}_{(i)}$ and integrated by parts in order to reduce the invariants to a canonical basis. Total derivatives, denoted by ``t.d.'', are neglected. The explicit results for the coefficients $C_{0}$-$C_{48f}$ in terms of the Galileon coupling $c_{2}$-$c_5$ and the Q-functional \eqref{QOrig} are provided in Appendix \ref{AppA}.
Comparing the result \eqref{FinalTraceT8} with the ansatz \eqref{AEAGalExp}, we find that only invariants with a higher number of derivatives per field than originally present in \eqref{AEAGalExp} are induced in the functional trace \eqref{FinalTraceT8}, implying
\begin{align}
\beta_{c_i}=0,\qquad i=2,\ldots,5.\label{betac}
\end{align}
This result is a consequence of the Galileon symmetry \eqref{Gsym} responsible for structure of the effective inverse Galileon metric \eqref{GInvExpl} and the operator $\Pi$ defined in \eqref{PiOp}. Since the $\left(G^{-1}_{i}\right)^{\mu\nu}$ only involve background tensor structures with two derivatives per field and the derivative expansion ultimately reduces to contractions among the $\left(G^{-1}_{i}\right)^{\mu\nu}$, projecting \eqref{FinalTraceT8} to the basis of Galileon invariants in \eqref{AEAGalExp} shows that there are no quantum induced contributions to the Galileon beta functions.
\subsection{Vertex expansion in momentum space}
In this section, we extract the beta functions by a vertex expansion in momentum space.
The vertex expansion involves the scale-dependent $n$-point functions
\begin{align}
\Gamma^{(n)}_k(x_1,\ldots,x_n):=\frac{\delta^n \Gamma_k}{\delta\varphi(x_1),\delta\varphi(x_2)\ldots\delta\varphi(x_n)}.\label{npoint}
\end{align}
Due to the scalar nature of the field $\varphi(x)$, the $n$-point functions \eqref{npoint} are totally symmetric under exchange of the $x_i$, $i=1,\ldots,n$.
Diagrammatically, the regulated Green's function $G_{\mathfrak{R}}(x,y)$ defined in \eqref{RegProp} is represented by a line.
\begin{figure}[h!]
	\begin{center}
		\begin{tikzpicture}
		\begin{scope}
		\draw[](0,0)--(3,0);
		\node at (-0.2,0) {$x$};
		\node at (3.2,0) {$y$};
		\node at (1.5,-0.3) {$G_{\mathfrak{R}}(x,y)$};
		\end{scope}
		\end{tikzpicture}
	\end{center}
	\caption{Diagrammatic representation of the regulated Greens function \eqref{RegProp}.}
	\label{Fig:Green}
\end{figure}

\noindent As shown in Fig.~\ref{Fig:Wett}, the functional trace is diagrammatically represented by joining the two ends (identifying $x=y$) of the line in Fig.~\ref{Fig:Green}, illustrating the one-loop structure of \eqref{Wetteq}.
\begin{figure}[h!]
	\begin{center}
		\begin{tikzpicture}
		\draw (0,0) circle (1cm);
		\begin{scope}
		\filldraw[white] (0,1) circle (0.3cm);
		\draw (0,1) circle (0.3cm);
		\draw[] (-0.25,1.2)--(0.25,0.8);
		\draw[] (0.25,1.2)--(-0.25,0.8);
		\end{scope}
		\node at (-1.9,0) {$\dot{\Gamma}_k\;\;=\frac{1}{2}\;\;$};
		\end{tikzpicture}
	\end{center}
	\caption{Diagrammatic representation of the Wetterich equation \eqref{Wetteq}. The encircled cross corresponds to an insertion of the scale derivative of the regulator $\otimes=k\partial_k\mathfrak{R}=\dot{\mathfrak{R}}$.}
	\label{Fig:Wett}
\end{figure}

\noindent Taking $n$ functional derivatives of the Wetterich equation \eqref{Wetteq} with respect to the Galileon field $\pi$ defines the functional flow of the $n$-point functions $\dot{\Gamma}^{(n)}_k$. The flow equations of the $\dot{\Gamma}^{(n)}_k$ are diagrammatically represented by insertions of $n$-point vertices $\Gamma^{(n)}_k$ into the loop.

In order to extract the beta functions for the Galileon truncation \eqref{AEAGalExp} from a vertex expansion, we need to derive the flow of the first five $n$-point functions, which are diagrammatically represented in Fig.~\ref{Fig:FlownPoint}. In general, the one-loop structure of the flow equation implies a hierarchy between the flow of the $n$-point functions, as $\dot{\Gamma}^{(n)}_k$ involves all $\Gamma^{(m)}_k$ with $m\leq n+2$.
\begin{figure}[h!]
	\begin{center}
		\begin{tikzpicture}[scale=0.4]
		\begin{scope}
		\draw (0,0) circle (1cm);
		\filldraw (-1,0) circle (2pt);
		\draw (-1.7,0)--(-1,0);
		\begin{scope}
		\filldraw[white] (0,1) circle (0.3cm);
		\node at (0,-1.3) {\tiny$(1a)$};
		\draw (0,1) circle (0.3cm);
		\draw[] (-0.25,1.2)--(0.25,0.8);
		\draw[] (0.25,1.2)--(-0.25,0.8);
		\end{scope}
		\node at (-4,0) {\small$\dot{\Gamma}^{(1)}_k\;\;=-\frac{1}{2}$};
		\end{scope}
		
		\begin{scope}[yshift=-3cm]
		\begin{scope}
		\draw (0,0) circle (1cm);
		\filldraw (-1,0) circle (2pt);
		\draw (-1.7,0)--(-1,0);
		\filldraw (1,0) circle (2pt);
		\draw (1.7,0)--(1,0);
		\begin{scope}
		\filldraw[white] (0,1) circle (0.3cm);
		\node at (0,-1.3) {\tiny$(2a)$};
		\draw (0,1) circle (0.3cm);
		\draw[] (-0.25,1.2)--(0.25,0.8);
		\draw[] (0.25,1.2)--(-0.25,0.8);
		\end{scope}
		\node at (-4.4,0) {\small $\dot{\Gamma}^{(2)}_k\;\;=\;\;$};
		\end{scope}
		\node at (2.5,0) {\small $-\frac{1}{2}$};
		\begin{scope}[xshift=4.6cm]
		\filldraw (0:1) circle (2pt);
		\draw (0:1)--(30:1.6);
		\draw (0:1)--(-30:1.6);
		\draw (0,0) circle (1cm);
		\begin{scope}
		\filldraw[white] (0,1) circle (0.3cm);
		\node at (0,-1.3) {\tiny$(2b)$};
		\draw (0,1) circle (0.3cm);
		\draw[] (-0.25,1.2)--(0.25,0.8);
		\draw[] (0.25,1.2)--(-0.25,0.8);
		\end{scope}
		\end{scope}
		\end{scope}
		
		\begin{scope}[yshift=-9cm]
		\node at (-4.4,0) {$\dot{\Gamma}^{(4)}_k\;\;=\;\;$};
		\node at (-2.4,0) {\small$12$};
		\begin{scope}
		\draw (0,0) circle (1cm);
		\filldraw (30:1) circle (2pt);
		\draw (30:1)--(30:1.6);
		\filldraw (-30:1) circle (2pt);
		\draw (-30:1)--(-30:1.6);
		\filldraw (30:-1) circle (2pt);
		\draw (30:-1)--(30:-1.6);
		\filldraw (-30:-1) circle (2pt);
		\draw (-30:-1)--(-30:-1.6);
		\begin{scope}
		\filldraw[white] (0,1) circle (0.3cm);
		\node at (0,-1.3) {\tiny$(4a)$};
		\draw (0,1) circle (0.3cm);
		\draw[] (-0.25,1.2)--(0.25,0.8);
		\draw[] (0.25,1.2)--(-0.25,0.8);
		\end{scope}
		\end{scope}
		
		\node at (2.4,0) {\small$-18$};
		
		\begin{scope}[xshift=4.6cm]
		\draw (0,0) circle (1cm);
		\filldraw (0:1) circle (2pt);
		\draw (0:1)--(30:1.6);
		\draw (0:1)--(-30:1.6);
		\filldraw (30:-1) circle (2pt);
		\draw (30:-1)--(30:-1.6);
		\filldraw (-30:-1) circle (2pt);
		\draw (-30:-1)--(-30:-1.6);
		\begin{scope}
		\filldraw[white] (0,1) circle (0.3cm);
		\node at (0,-1.3) {\tiny$(4b)$};
		\draw (0,1) circle (0.3cm);
		\draw[] (-0.25,1.2)--(0.25,0.8);
		\draw[] (0.25,1.2)--(-0.25,0.8);
		\end{scope}
		\end{scope}
		
		\node at (6.7,0) {\small $+3$};

		\begin{scope}[xshift=8.6cm]
		\draw (0,0) circle (1cm);
		\filldraw (0:1) circle (2pt);
		\draw (0:1)--(30:1.6);
		\draw (0:1)--(-30:1.6);
		\filldraw (0:-1) circle (2pt);
		\draw (0:-1)--(30:-1.6);
		\draw (0:-1)--(-30:-1.6);
		\begin{scope}
		\filldraw[white] (0,1) circle (0.3cm);
		\node at (0,-1.3) {\tiny$(4c)$};
		\draw (0,1) circle (0.3cm);
		\draw[] (-0.25,1.2)--(0.25,0.8);
		\draw[] (0.25,1.2)--(-0.25,0.8);
		\end{scope}
		\end{scope}
		\end{scope}

		\begin{scope}[yshift=-9cm]
		\begin{scope}[xshift=12.9cm]
		\node at (-2.3,0) {\small$+4$};
		\draw (0,0) circle (1cm);
		\filldraw (0:1) circle (2pt);
		\draw (0:1)--(30:1.6);
		\draw (0:1)--(-30:1.6);
		\draw (0:1)--(1.4,0);
		\filldraw (-1,0) circle (2pt);
		\draw (-1.6,0)--(-1,0);
		\begin{scope}
		\filldraw[white] (0,1) circle (0.3cm);
		\node at (0,-1.3) {\tiny$(4d)$};
		\draw (0,1) circle (0.3cm);
		\draw[] (-0.25,1.2)--(0.25,0.8);
		\draw[] (0.25,1.2)--(-0.25,0.8);
		\end{scope}
		\end{scope}
		
		\begin{scope}[xshift=17.3cm]
		\node at (-2,0) {\small $-\frac{1}{2}$};
		\filldraw (0:1) circle (2pt);
		\draw (0:1)--(30:1.6);
		\draw (0:1)--(-30:1.6);
		\draw (0:1)--(15:1.6);
		\draw (0:1)--(-15:1.6);
		\draw (0,0) circle (1cm);
		\begin{scope}
		\filldraw[white] (0,1) circle (0.3cm);
		\node at (0,-1.3) {\tiny$(4e)$};
		\draw (0,1) circle (0.3cm);
		\draw[] (-0.25,1.2)--(0.25,0.8);
		\draw[] (0.25,1.2)--(-0.25,0.8);
		\end{scope}
		\end{scope}
		
		\end{scope}
		
		\begin{scope}[yshift=-6cm]
		\node at (-4.4,0) {$\dot{\Gamma}^{(3)}_k\;\;=\;\;$};
		\node at (-2.4,0) {\small$-3$};
		\begin{scope}
		\draw (0,0) circle (1cm);
		\filldraw (30:1) circle (2pt);
		\draw (30:1)--(30:1.6);
		\filldraw (-30:1) circle (2pt);
		\draw (-30:1)--(-30:1.6);
		\filldraw (-1,0) circle (2pt);
		\draw (-1.6,0)--(-1,0);
		\begin{scope}
		\filldraw[white] (0,1) circle (0.3cm);
		\node at (0,-1.3) {\tiny$(3a)$};
		\draw (0,1) circle (0.3cm);
		\draw[] (-0.25,1.2)--(0.25,0.8);
		\draw[] (0.25,1.2)--(-0.25,0.8);
		\end{scope}
		\end{scope}
		
		\node at (2.4,0) {\small$+3$};
		
		\begin{scope}[xshift=4.6cm]
		\draw (0,0) circle (1cm);
		\filldraw (0:1) circle (2pt);
		\draw (0:1)--(30:1.6);
		\draw (0:1)--(-30:1.6);
		\filldraw (-1,0) circle (2pt);
		\draw (-1.6,0)--(-1,0);
		\begin{scope}
		\filldraw[white] (0,1) circle (0.3cm);
		\node at (0,-1.3) {\tiny$(3b)$};
		\draw (0,1) circle (0.3cm);
		\draw[] (-0.25,1.2)--(0.25,0.8);
		\draw[] (0.25,1.2)--(-0.25,0.8);
		\end{scope}
		\end{scope}
		
		\node at (6.7,0) {\small$-\frac{1}{2}$};
		
		\begin{scope}[xshift=8.6cm]
		\filldraw (0:1) circle (2pt);
		\draw (0:1)--(30:1.6);
		\draw (0:1)--(-30:1.6);
		\draw (0:1)--(1.4,0);
		\draw (0,0) circle (1cm);
		\begin{scope}
		\filldraw[white] (0,1) circle (0.3cm);
		\node at (0,-1.3) {\tiny$(3c)$};
		\draw (0,1) circle (0.3cm);
		\draw[] (-0.25,1.2)--(0.25,0.8);
		\draw[] (0.25,1.2)--(-0.25,0.8);
		\end{scope}
		\end{scope}
		\end{scope}
		
	\begin{scope}[yshift=-12cm]
\node at (-4.4,0) {$\dot{\Gamma}^{(5)}_k\;\;=\;\;$};
\node at (-2.3,0) {\small$-60$};
\begin{scope}
\draw (0,0) circle (1cm);
\filldraw (30:1) circle (2pt);
\draw (30:1)--(30:1.6);
\filldraw (-30:1) circle (2pt);
\draw (-30:1)--(-30:1.6);
\filldraw (40:-1) circle (2pt);
\draw (40:-1)--(40:-1.6);
\filldraw (0:-1) circle (2pt);
\draw (0:-1)--(0:-1.4);
\filldraw (-40:-1) circle (2pt);
\draw (-40:-1)--(-40:-1.6);
\begin{scope}
\filldraw[white] (0,1) circle (0.3cm);
\node at (0,-1.3) {\tiny$(5a)$};
\draw (0,1) circle (0.3cm);
\draw[] (-0.25,1.2)--(0.25,0.8);
\draw[] (0.25,1.2)--(-0.25,0.8);
\end{scope}
\end{scope}

\node at (2.1,0) {\small$+120$};

\begin{scope}[xshift=4.6cm]
\draw (0,0) circle (1cm);
\filldraw (0:1) circle (2pt);
\draw (0:1)--(30:1.6);
\draw (0:1)--(-30:1.6);
\filldraw (40:-1) circle (2pt);
\draw (40:-1)--(40:-1.6);
\filldraw (0:-1) circle (2pt);
\draw (0:-1)--(0:-1.4);
\filldraw (-40:-1) circle (2pt);
\draw (-40:-1)--(-40:-1.6);
\begin{scope}
\filldraw[white] (0,1) circle (0.3cm);
\node at (0,-1.3) {\tiny$(5b)$};
\draw (0,1) circle (0.3cm);
\draw[] (-0.25,1.2)--(0.25,0.8);
\draw[] (0.25,1.2)--(-0.25,0.8);
\end{scope}
\end{scope}

\node at (6.6,0) {\small$-45$};

\begin{scope}[xshift=8.6cm]
\draw (0,0) circle (1cm);
\filldraw (0:1) circle (2pt);
\draw (0:1)--(30:1.6);
\draw (0:1)--(-30:1.6);
\filldraw (30:-1) circle (2pt);
\draw (30:-1)--(40:-1.6);
\draw (30:-1)--(20:-1.6);
\filldraw (-30:-1) circle (2pt);
\draw (-30:-1)--(-30:-1.6);
\begin{scope}
\filldraw[white] (0,1) circle (0.3cm);
\node at (0,-1.3) {\tiny$(5c)$};
\draw (0,1) circle (0.3cm);
\draw[] (-0.25,1.2)--(0.25,0.8);
\draw[] (0.25,1.2)--(-0.25,0.8);
\end{scope}
\end{scope}
\end{scope}

\begin{scope}[yshift=-12cm]
\begin{scope}[xshift=12.9cm]
\node at (-2.2,0) {\small $-30$};
\draw (0,0) circle (1cm);
\filldraw (0:1) circle (2pt);
\draw (0:1)--(30:1.6);
\draw (0:1)--(-30:1.6);
\draw (0:1)--(1.4,0);
\filldraw (30:-1) circle (2pt);
\draw (30:-1)--(30:-1.6);
\filldraw (-30:-1) circle (2pt);
\draw (-30:-1)--(-30:-1.6);
\begin{scope}
\filldraw[white] (0,1) circle (0.3cm);
\node at (0,-1.3) {\tiny$(5d)$};
\draw (0,1) circle (0.3cm);
\draw[] (-0.25,1.2)--(0.25,0.8);
\draw[] (0.25,1.2)--(-0.25,0.8);
\end{scope}
\end{scope}

\begin{scope}[xshift=17.2cm]
\node at (-2,0) {\small$+10$};
\draw (0,0) circle (1cm);
\filldraw (0:1) circle (2pt);
\draw (0:1)--(30:1.6);
\draw (0:1)--(-30:1.6);
\draw (0:1)--(1.4,0);
\filldraw (0:-1) circle (2pt);
\draw (0:-1)--(30:-1.6);
\draw (0:-1)--(-30:-1.6);
\begin{scope}
\filldraw[white] (0,1) circle (0.3cm);
\node at (0,-1.3) {\tiny$(5e)$};
\draw (0,1) circle (0.3cm);
\draw[] (-0.25,1.2)--(0.25,0.8);
\draw[] (0.25,1.2)--(-0.25,0.8);
\end{scope}
\end{scope}

\begin{scope}[xshift=21.4cm]
\node at (-2.2,0) {\small$+5$};
\draw (0,0) circle (1cm);
\filldraw (0:1) circle (2pt);
\draw (0:1)--(30:1.6);
\draw (0:1)--(-30:1.6);
\draw (0:1)--(15:1.6);
\draw (0:1)--(-15:1.6);
\draw (0,0) circle (1cm);
\filldraw (0:-1) circle (2pt);
\draw (0:-1)--(0:-1.6);
\begin{scope}
\filldraw[white] (0,1) circle (0.3cm);
\node at (0,-1.3) {\tiny$(5f)$};
\draw (0,1) circle (0.3cm);
\draw[] (-0.25,1.2)--(0.25,0.8);
\draw[] (0.25,1.2)--(-0.25,0.8);
\end{scope}
\end{scope}

\begin{scope}[xshift=25.7cm]
\node at (-2,0) {\small$-\frac{1}{2}$};
\draw (0,0) circle (1cm);
\filldraw (0:1) circle (2pt);
\draw (0:1)--(30:1.6);
\draw (0:1)--(-30:1.6);
\draw (0:1)--(0:1.6);
\draw (0:1)--(15:1.6);
\draw (0:1)--(-15:1.6);
\draw (0,0) circle (1cm);
\begin{scope}
\filldraw[white] (0,1) circle (0.3cm);
\node at (0,-1.3) {\tiny$(5g)$};
\draw (0,1) circle (0.3cm);
\draw[] (-0.25,1.2)--(0.25,0.8);
\draw[] (0.25,1.2)--(-0.25,0.8);
\end{scope}
\end{scope}

\end{scope}
		\end{tikzpicture}
	\end{center}
	\caption{Diagrammatic representation of the RG flow $\dot{\Gamma}_k^{(n)}$ of the $n$-point functions $\Gamma_k^{(n)}$. The black dots represent insertions of $n$-point vertices $\Gamma^{(n)}_k$, the solid lines represent regulated propagators $G_{\mathfrak{R}}$. The numerical coefficient in front of each diagram results from the functional differentiation of \eqref{Wetteq}.}
\label{Fig:FlownPoint}
\end{figure}

\noindent Evaluating the functional derivatives \eqref{npoint} at $\pi=0$ and performing a Fourier transform,
we obtain the momentum space expressions for the regulated propagator $G_{\mathfrak{R}}(p)=1/\Gamma^{(2)}(p_1,-p_1)$ and the interaction vertices $\Gamma^{(n)}(p_1,\ldots,p_n)$ with $n>2$.
For the inner product between two four-momentum vectors $p_{\mu}^{i}$ and $p_{\mu}^{j}$ we write
\begin{align}
(p_i\cdot p_j):=p^{i}_{\mu}\,p^{j}_{\nu}\,\delta^{\mu\nu},\qquad p_i^2:=(p_i\cdot p_i).
\end{align}
Using the convention that all momenta are incoming and stripping off an overall momentum-conservating delta function $\delta(\sum_{i=1}^{n} p^{i}_{\mu})$, we obtain   
\begin{align}
G_{\mathfrak{R}}(p)={}&\frac{1}{c_2p^2+\mathfrak{R}(p^2)},\label{PropMom}\\
\Gamma^{(n)}_k(p_1,\ldots,p_n)={}&c_{n}\frac{(n-1)!}{2n}\mathfrak{M}(p_1,\ldots,p_n),\qquad n>2\label{vertex}.
\end{align}
The,
 $\mathfrak{M}(p_1,\ldots,p_n)$ are defined as the sum of the diagonal minors $\mathfrak{M}_{(ii)}(p_1,\ldots,\hat{p}_i,\ldots p_n)$ of the ${n\times n}$ Gram matrix $\mathfrak{G}_{ij}:=(p_i\cdot p_j)$.\footnote{That is, $\mathfrak{M}(p_1,\ldots,p_n)=\sum_{i=1}^{n}\mathfrak{M}_{(ii)}(p_1,\ldots,\hat{p}_i,\ldots p_n)$ where $\mathfrak{M}_{(ii)}(p_1,\ldots,\hat{p}_i,\ldots p_n)$ is the determinant of the matrix obtained by deleting the $i$th row and $i$th column of $\mathfrak{G}_{ij}$.}
Despite the fact that we are working in $d=4$, in order to make the $d$-dependent numerical factors more transparent, we write the analytic expressions of the $\dot{\Gamma}^{(n)}_k$ corresponding to the one-loop diagrams in Fig.~\ref{Fig:FlownPoint} for general $d$ and reduce the expressions to $d=4$ at the end. For example, the RG running of the two-point functions reads
\begin{align}
\dot{\Gamma}^{(2)}(p,-p)={}&\int\frac{\mathrm{d}\ell^d}{(2\pi)^d}\left[G(\ell)\,\Gamma^{(3)}(\ell,p,-p-\ell)\, G_{\mathfrak{R}}(p+\ell)\, \Gamma^{(3)}(p+\ell,-p,-\ell)\, G_{\mathfrak{R}}(\ell)\, \dot{\mathfrak{R}}(\ell)\right]\nonumber\\
&-\frac{1}{2}\int\frac{\mathrm{d}\ell^d}{(2\pi)^d}\left[G_{\mathfrak{R}}(\ell)\,\Gamma^{(4)}(\ell,p,-p,-\ell)\,G_{\mathfrak{R}}(\ell)\, \dot{\mathfrak{R}}(l)\right].\label{flow2p}
\end{align}
We refrain from providing the analytic expressions for the momentum integrals of the diagrams that enter the flow equation of the higher $n$-point functions, since the general structure and the momentum flow is analogue to that in \eqref{flow2p}. Momentum conservation requires that the sum of all momenta at each vertex add up to zero and overall momentum conservation $\sum_{i=1}^{n}p_{\mu}^{i}=0$ permits to express the $n$ external momenta $p_{\mu}^{i}$, $i=1,\ldots,n$ through a linear combination of $n-1$ external momenta.
 
Looking at the structure of the Galileon truncation \eqref{Gact}, the beta functions of the Galileon operators are obtained by extracting those parts of the $\dot{\Gamma}^{(n)}_k$ that have $2(n-1)$ powers of the external momenta. The vertices $\Gamma^{(n)}_k$ only contribute positive powers of external momenta. Expanding the propagators around zero external momenta up to the required order, all contributions with a particular power of external momenta contributing to the $\dot{\Gamma}^{(n)}_k$ can be extracted.
For example, the expansion of the propagator up to $\mathcal{O}\left(p^4\right)$ reads 
\begin{align}
G_{\mathfrak{R}}(\ell+p)={}&G_{\mathfrak{R}}(\ell)+2G^{(1)}_{\mathfrak{R}}(\ell)(\ell\cdot p)+\left[2G^{(2)}_{\mathfrak{R}}(\ell)(\ell\cdot p)^2+G^{(1)}_{\mathfrak{R}}(\ell)\,p^2\right]\nonumber\\
&+\frac{1}{3}\left[4G^{(3)}_{\mathfrak{R}}(\ell)(\ell\cdot p)^3+6G^{(2)}_{\mathfrak{R}}(\ell)(\ell\cdot p)\,p^2\right]\nonumber\\
&+\frac{1}{6}\left[4G^{(4)}_{\mathfrak{R}}(\ell)(\ell\cdot p)^4+12G^{(3)}_{\mathfrak{R}}(\ell)(\ell\cdot p)^2\,p^2+3G^{(2)}_{\mathfrak{R}}(\ell)\,p^4\right]+\mathcal{O}(p^5).\label{PropExp}
\end{align}
The superscript denotes derivatives with respect to the argument $G^{(n)}_{\mathfrak{R}}(\ell)=\partial^nG_{\mathfrak{R}}(z)/\partial z^n|_{z=\ell^2}$.
The tensor loop integrals are reduced to scalar ones by exploiting Lorentz invariance \cite{Steinwachs2019},
\begin{align}
\int\mathrm{d}^d\ell\, f^{\mu_1\ldots\mu_{2n}}(p)\ell_{\mu_1}\ldots\ell_{\mu_{2n}}=\frac{\Gamma(d/2)}{2^n\Gamma(n+d/2)}\int\mathrm{d}^d\ell \left(\ell^2\right)^{n}\, f^{\mu_1\ldots\mu_{2n}}(p)\,[\delta_{\mathrm{sym}}^{n}]_{\mu_1\ldots\mu_{2n}}.\label{lint}
\end{align}
Here, $[\delta_{\mathrm{sym}}^{n}]_{\mu_1\ldots\mu_{2n}}$ is defined in \eqref{dsym} and the prefactor in \eqref{lint} is determined by the trace
\begin{align}
[\delta_{\mathrm{sym}}^{n}]_{\mu_1\ldots\mu_{2n}}\delta^{\mu_1\mu_2}\ldots\delta^{\mu_{2n-1}\mu_{2n}}=\frac{2^n\Gamma(n+d/2)}{\Gamma(d/2)}.
\end{align}
Transforming to polar loop coordinates and performing the angular integrals results in 
\begin{align}
\int\frac{\mathrm{d}^d\ell}{(2\pi)^d}=\frac{S_d}{(2\pi)^d}\int\mathrm{d}\ell\,\ell^{d-1}.
\end{align}
The surface of the unit sphere in $d$ dimensions is defined by $S_d:=2\pi^{d/2}/\Gamma(d/2)$.
Finally, we transform the loop variable $z=\ell^2$, $\mathrm{d}z=2\ell\mathrm{d}\ell$ and use the definition \eqref{QOrig} in order to absorb the remaining loop integrals in the Q-functionals.
In order to extract the beta functions of the Galileon couplings, we extract those contributions to the $\dot{\Gamma}^{(n)}_k$ with $2(n-1)$ powers of external momenta and find
\begin{align}
\left.\dot{\Gamma}^{(n)}(p_1,\ldots,p_n)\right|_{p^{2(n-1)}}=0.
\end{align}   
In agreement with the result \eqref{FinalTraceT8}, this implies that all beta functions vanish
\begin{align}
\beta_{c_i}=0,\qquad i=2,3,4,5.\label{ZeroBeta}
\end{align}
The result \eqref{ZeroBeta} can again directly be traced back to the Galileon symmetry \eqref{Gsym} and the one-loop structure of the Wetterich equation \eqref{Wetteq}. The Galileon symmetry implies that each $n$-point vertex $\Gamma^{(n)}_k$ carries $2(n-1)$ powers of momenta. The one-loop structure implies that each vertex $\Gamma^{(n)}_k$ inserted in the one-loop diagrams in Fig.~\ref{Fig:FlownPoint} will at least contribute $2(n-2)$ powers of external momenta, since at most two legs can carry a loop momentum.\footnote{In addition, the $6$-point and $7$-point vertices entering the diagrams $(4c)$, $(5f)$ and $(5g)$ in Fig.~\ref{Fig:FlownPoint} are trivially zero in the Galileon truncation \eqref{Gact}.} In view of the expansion \eqref{PropExp}, each propagator can only further increase the power of external momenta in a diagram.
Hence, all $n$-point one-loop diagrams in Fig.~\ref{Fig:FlownPoint} have at least $2n>2(n-1)$ powers of external momenta and therefore cannot contribute to the Galileon beta functions.
\subsection{Covariant geometric resummation}
\label{CovGeoSum}
The effective Galileon metric \eqref{Gmetric} defines the required geometric structure for an application of covariant heat kernel techniques for the fluctuation operator \eqref{Op}. Following the approach proposed in the context of the perturbative one-loop calculation \cite{Heisenberg:2019wjv}, 
We define $\nabla^{\mathcal{G}}_{\mu}$ as the torsion-free covariant derivative compatible with $\mathcal{G}_{\mu\nu}$,
\begin{align}
[\nabla^{\mathcal{G}}_{\mu},\nabla^{\mathcal{G}}_{\nu}]\pi=0,\qquad
\nabla^{\mathcal{G}}_{\rho}\mathcal{G}_{\mu\nu}=0,
\end{align}
The connection $\tensor{\Gamma}{^{\rho}_{\mu\nu}}(\mathcal{G})$ associated to $\nabla^{\mathcal{G}}$ reads
\begin{align}
\tensor{\Gamma}{^{\rho}_{\mu\nu}}(\mathcal{G})=\frac{1}{2}\left(\mathcal{G}^{-1}\right)^{\rho\sigma}\left(\partial_{\mu}\mathcal{G}_{\sigma\nu}+\partial_{\nu}\mathcal{G}_{\sigma\mu}-\partial_{\sigma}\mathcal{G}_{\mu\nu}\right).\label{GGam}
\end{align}
Indices are raised and lowered exclusively with $\mathcal{G}_{\mu\nu}$ and $\left(\mathcal{G}^{-1}\right)^{\mu\nu}$, respectively. We define the positive definite  covariant Laplacian as
\begin{align}
\Delta_{\mathcal{G}}:=-\left(\mathcal{G}^{-1}\right)^{\mu\nu}\nabla^{\mathcal{G}}_{\mu}\nabla^{\mathcal{G}}_{\nu}.\label{GLap}
\end{align}
When the Laplacian $\Delta_{\mathcal{G}}$ acts on scalars, it is related to the fluctuation operator \eqref{Op} by
\begin{align}
F(\partial)=-\left(\mathcal{G}^{-1}\right)^{\mu\nu}\partial_{\mu}\partial_{\nu}=\Delta_{\mathcal{G}}-\left(\mathcal{G}^{-1}\right)^{\mu\nu}\tensor{\Gamma}{^{\rho}_{\mu\nu}}(\mathcal{G})\nabla^{\mathcal{G}}_{\rho}.\label{LapG}
\end{align}
In addition, we define the ``bundle connection'' acting on scalars
\begin{align}
\Sigma^{\rho}:=\frac{1}{2}\left(\mathcal{G}^{-1}\right)^{\mu\nu}\tensor{\Gamma}{^{\rho}_{\mu\nu}}(\mathcal{G})
={}&\det(\mathcal{G})^{1/4}\left(\mathcal{G}^{-1}\right)^{\rho\mu}\partial_{\mu}\det(\mathcal{G})^{-1/4}.\label{GCon}
\end{align} 
Combining \eqref{LapG} with \eqref{GCon}, the operator \eqref{GLap} can be written as
\begin{align}
F(\nabla^{\mathcal{G}})=\Delta_{\mathcal{G}}-2\Sigma^{\rho}\nabla_{\rho}^{\mathcal{G}}.\label{Op2}
\end{align}
In terms of $\mathcal{D}_{\mu}:=\nabla^{\mathcal{G}}_{\mu}+\mathcal{G}_{\mu\nu}\Sigma^{\nu}$, the operator \eqref{Op2} acquires minimal second order form
\begin{align}
F(\mathcal{D})=-\mathcal{D}^2+E,\label{Op3}
\end{align}
with $-\mathcal{D}^2=-\left(\mathcal{G}^{-1}\right)^{\mu\nu}\mathcal{D}_{\mu}\mathcal{D}_{\mu}$ and the endomorphism
\begin{align}
E:=\nabla_{\nu}^{G}\Sigma^{\nu}+\Sigma_{\nu}\Sigma^{\nu}.\label{PotP}
\end{align}
For the scalar $\pi$, the bundle curvature vanishes due to the antisymmetry of the commutator,
\begin{align}
\mathcal{R}_{\mu\nu}\pi:=[\mathcal{D}_{\mu},\mathcal{D}_{\nu}]\pi=0.
\end{align}
Adopting a ``spectrally adjusted'' type III regulator \cite{Codello2009}, for which the argument is identified with the minimal second-order operator \eqref{Op3}, we obtain the regulated Greens function
\begin{align}
G_{\mathfrak{R}}=\frac{1}{P(-\mathcal{D}^2+E)},\qquad P(-\mathcal{D}^2+E):=-\mathcal{D}^2+E+\mathfrak{R}(-\mathcal{D}^2+E).\label{GreenMinOp}
\end{align}
In terms of this geometric reformulation, an infinite number of operators are resummed into curvatures of the effective Galileon metric $\mathcal{G}_{\mu\nu}$ which have the schematic structure ${R(\mathcal{G})\sim\partial (\mathcal{G}^{-1}\partial \mathcal{G})+(\mathcal{G}^{-1}\partial G)^2}$. Since $\mathcal{G}=\mathcal{O}(\partial^2)$, it follows $R_{\mu\nu\rho\sigma}(\mathcal{G})=\mathcal{O}\left(\partial^4\right)$ such that an expansion up to BDO $\mathcal{O}(\partial^8)$ corresponds to an expansion up to $\mathcal{O}(R^2)$.
The functional trace \eqref{TTrace} with the Greens function \eqref{GreenMinOp} of the covariant operator \eqref{Op3} reads
\begin{align}
T={}&\frac{1}{2}\mathrm{Tr}\left(G_{\mathfrak{R}}\,\dot{\mathfrak{R}}\right)={}\frac{1}{2}\int\mathrm{d}^4x\int\mathrm{d}s\mathcal{L}^{-1}\left[G_{\mathfrak{R}}\,\dot{\mathfrak{R}}\right](s)\,\langle x| e^{-s\left(-\mathcal{D}^2+E\right)}|x\rangle\nonumber\\
={}&\frac{1}{2}\int\mathrm{d}^4x\sqrt{\mathcal{G}}\int\frac{\mathrm{d}s}{(4\pi s)^{2}}\mathcal{L}^{-1}\left[G_{\mathfrak{R}}\,\dot{\mathfrak{R}}\right](s)\left[1+s\,a_1+s^2a_2+\mathcal{O}\left(s^3\right)\right]\nonumber\\
={}&\frac{1}{2}\int\mathrm{d}^4x\sqrt{\mathcal{G}}\left\{Q_2[G_{\mathfrak{R}}\,\dot{\mathfrak{R}}]+2Q_{1}\left[G_{\mathfrak{R}}\,\dot{\mathfrak{R}}\right]a_1+4Q_0\left[G_{\mathfrak{R}}\,\dot{\mathfrak{R}}\right]a_2+\mathcal{O}\left(R^3\right)\right\}.\label{TraceGeometric}
\end{align}
Here, $a_1$ and $a_2$ are the traced coincidence limits of the first two Schwinger-DeWitt coefficients for the minimal scalar second-order operator \eqref{Op3},
\begin{align}
a_1(x,x)={}&E-\frac{1}{6}R(\mathcal{G}),\\
a_2(x,x)={}&\frac{1}{180}\left[R_{\mu\nu\rho\sigma}(\mathcal{G})R^{\mu\nu\rho\sigma}(\mathcal{G})-R_{\mu\nu}(\mathcal{G})R^{\mu\nu}(\mathcal{G})\right]+\frac{1}{2}\left(E-\frac{1}{6}R(\mathcal{G})\right)^2+\mathrm{t.d.}
\end{align}
Similar to the procedure in \cite{Heisenberg2019}, the result of the functional trace in terms of the original Galileon field is recovered by expanding \eqref{TraceGeometric} in powers of perturbations $\delta^n_\pi\left(\mathcal{G}^{-1}\right)^{\mu\nu}$ evaluated at zero mean field $\pi$. In view of the explicit expression \eqref{GInvExpl} we find the structural relations
\begin{align}
\delta_{\pi}\left(\mathcal{G}^{-1}\right)|_{\pi=0}{}&\sim\partial^2\delta\pi,\label{s1}\\
\delta^2_{\pi}\left(\mathcal{G}^{-1}\right)|_{\pi=0}{}&\sim\left(\partial^2\delta\pi\right)\left(\partial^2\delta\pi\right),\\
\delta^3_{\pi}\left(\mathcal{G}^{-1}\right)|_{\pi=0}{}&\sim\left(\partial^2\delta\pi\right)\left(\partial^2\delta\pi\right)\left(\partial^2\delta\pi\right),\\
\delta^n_\pi\left(\mathcal{G}^{-1}\right)|_{\pi=0}{}&=0\label{s3},\qquad n>3.
\end{align}
It is easy to see that the derivative structure of the invariants obtained by expanding the resummed geometric result in powers of $\delta\pi$ coincides with the structures obtained in the derivative expansion \eqref{FinalTraceT8}. In particular, for all invariants generated by this expansion the number of derivatives per field is higher than for the invariants present in the Galileon truncation \eqref{AEAGalExp}.
For example, extracting the invariants that involving two powers of $\delta\pi$ leads to the structural relations,
\begin{align}
\delta^2_{\pi}\sqrt{\mathcal{G}}|_{\pi=0}{}&\sim\left[\delta_\pi\left(\mathcal{G}^{-1}\right)\delta_\pi\left(\mathcal{G}^{-1}\right)+\delta^2_\pi\left(\mathcal{G}^{-1}\right)\right]_{\pi=0}\sim\left(\partial^2\delta\pi\right)\left(\partial^2\delta\pi\right),\label{s11}\\
\delta^2_{\pi} \sqrt{\mathcal{G}}R(\mathcal{G})|_{\pi=0}{}&\sim\left[\delta_\pi\left(\mathcal{G}^{-1}\right)\partial^2\delta_\pi\left(\mathcal{G}^{-1}\right)+\partial^2\delta^2_\pi\left(\mathcal{G}^{-1}\right)\right]_{\pi=0}\sim\left(\partial^3\delta\pi\right)\left(\partial^3\delta\pi\right),\\
\delta^2_{\pi} \sqrt{\mathcal{G}}E(\mathcal{G})|_{\pi=0}{}&\sim\left[\delta_\pi\left(\mathcal{G}^{-1}\right)\partial^2\delta_\pi\left(\mathcal{G}^{-1}\right)+\partial^2\delta^2_\pi\left(\mathcal{G}^{-1}\right)\right]_{\pi=0}\sim\left(\partial^3\delta\pi\right)\left(\partial^3\delta\pi\right),\\
\delta^2_{\pi}\sqrt{\mathcal{G}}R^2(\mathcal{G})|_{\pi=0}{}&\sim\partial^2\left(\delta_\pi\mathcal{G}^{-1}\right)\partial^2\left(\delta_{\pi}\mathcal{G}^{-1}\right)|_{\pi=0}\sim\left(\partial^4\delta\pi\right)\left(\partial^4\delta\pi\right).\\
\delta^2_{\pi}\sqrt{\mathcal{G}}E^2(\mathcal{G})|_{\pi=0}{}&\sim\partial^2\left(\delta_\pi\mathcal{G}^{-1}\right)\partial^2\left(\delta_{\pi}\mathcal{G}^{-1}\right)|_{\pi=0}\sim\left(\partial^4\delta\pi\right)\left(\partial^4\delta\pi\right).\label{s15}
\end{align}
Hence, no invariant $\left(\delta\pi\right)\partial^2\left(\partial\pi\right)$ which is present in the ansatz \eqref{AEAGalExp} will be generated by the re-expanding  \eqref{TraceGeometric} in terms of the Galileon field. This pattern continues for invariants involving higher powers\footnote{Each derivative w.r.t. $\pi$ increases the number of fields by one. From the invariant line element $\sqrt{\mathcal{G}}$ only structures with two derivatives per field are generated (which follows from \eqref{s1}-\eqref{s3}), while additional curvature $R(\mathcal{G})$ or endomorphism $E(\mathcal{G})$ adds two additional derivatives.} of $\delta\pi$ and, in agreement with \eqref{betac} and \eqref{ZeroBeta}, implies  
\begin{align}
\beta_{c_i}=0,\qquad i=2,3,4,5.\label{betazero}
\end{align} 
Again the Galileon symmetry which leads to the particular structure of the inverse Galileon metric is responsible for the structural relations \eqref{s1}-\eqref{s3}, \eqref{s11}-\eqref{s15} and the vanishing beta functions \eqref{betazero}. 

\section{Galileon beta functions and fixed points}
\label{Discussion}

According to \eqref{MassDim}, the Galileon couplings $c_i$ have negative mass dimension. Since the RG scale $k$ has mass dimension $[k]=1$, we define the dimensionless Galileon couplings $\tilde{c}_i$ by rescaling the $c_i$ with the appropriate power of $k$,
\begin{align}
c_i:=k^{-3(i-2)}\tilde{c}_i.
\end{align}
The beta functions $\beta_{\tilde{c}_i}:=\dot{\tilde{c}}_i$ for the dimensional couplings $\tilde{c}_i$ are obtained by
\begin{align}
0=\beta_{c}=\dot{c}_i=k\partial_kc_i=\left[-3(i-2)\tilde{c}_i+\dot{\tilde{c}}_i\right]k^{-3(i-2)}.
\end{align}
Hence, the Galileon RG system does not receive any quantum contributions and the only contribution to the dimensionless Galileon beta functions arise form the canonical mass scaling of the couplings $c_i$, 
\begin{align}
\beta_{\tilde{c}_i}=3(i-2)\tilde{c}_i.\label{betadimless}
\end{align}
A fixed point $\tilde{c}_i^{*}$ is defined by $\beta_{\tilde{c}_i}(\tilde{c}_i^{*})=0$. Consequently, the the Galileon RG system \eqref{betadimless} only features the trivial Gaussian IR fixed point
\begin{align}
\tilde{c}_i^{*}=0,\qquad i=2,3,4,5.
\end{align}
This result is independent of the expansion scheme and choice of the regulator and a direct consequence of the Galileon symmetry \eqref{Gsym}.
\section{Conclusion}
\label{Conclusion}
We have investigated the non-perturbative RG flow of the scalar Galileon in flat space within the ERGE. We have evaluated the functional trace in the Wetterich equation by various expansion schemes with different type of regulators, including on a heat-kernel technique based $\mathcal{O}(\partial^8)$ derivative expansion, a vertex expansion in momentum space and a covariant formulation based on a geometric resummation of an infinite number of operators in terms of curvature tensors of an effective Galileon metric. 

Independently of the method and the type of regulator, we have found that the RG flow of the dimensionless Galileon couplings is solely driven by their classical mass dimension, that is the beta functions do not receive any quantum-induced contributions. This leads to the main result that the only fixed point of the Galileon truncation is the trivial Gaussian IR fixed point. 

This result might be natural to expect from and is in agreement with the non-renorma\-lization property of the Galileon found in the perturbative quantization  \cite{Luty:2003vm,Hinterbichler:2010xn,Goon:2016ihr,Heisenberg2019a, Heisenberg:2019wjv}. However, in contrast to the counterterms that arise in dimensional regulated perturbative calculations, the functional trace in the Wetterich equation also induces operators which correspond to power-law divergences in perturbative one-loop calculations (which are annihilated by dimensional regularization). Nevertheless, even these additional operators are not of the form of the operators present in the ansatz for the EAA \eqref{AEAGalExp} such that a projection of the functional trace to the Galileon truncation vanishes.  
As emphasized, the origin of this result is directly connected to the Galileon symmetry, which manifests itself in different ways in the derivative expansion, the momentum space vertex expansion and the geometric resummation

It would be interesting to investigate the non-perturbative RG flow of a general shift-symmetric higher-derivative scalar field theory which breaks the Galileon symmetry \eqref{Gsym} within a derivative expansion of the ERGE . This would allow to compare the structure of the RG flow for less symmetric theory with the enhanced Galileon symmetry \cite{Steinwachs2021}. 
Beside breaking the Galileon symmetry directly, a natural extension of the present work would be the non-perturbative RG flow of the covariant Galileon in curved space. Due to the presence of the gravitational interaction, the beta functions of the couplings in the resulting higher-derivative scalar-tensor theory will receive quantum contributions and might feature a non-trivial interacting UV fixed point underlying the asymptotic safety scenario. In this context, it would also be interesting to study the relation between the RG flow of this higher-derivative theories with that of scalar-tensor theories and geometric $f(R)$ theories, whose renormalization structure has been studied in perturbative one-loop calculations \cite{Barvinsky1993,Shapiro1995,Steinwachs2011,Steinwachs2012,Kamenshchik2015,Ruf2018,Ruf2018c} and the ERGE \cite{Machado2008,Codello2008,Narain2010, Narain:2009gb,Percacci2011,Henz:2013oxa,Benedetti:2013nya,Percacci2015,Labus:2015ska,Henz:2016aoh,Merzlikin2017,Martini:2018ska,Eichhorn:2020sbo}.

\acknowledgments
I thank Roberto Percacci for many interesting conversations and constructive comments on the first version of the draft. I also thank Omar Zanusso for helpful discussions. 
\appendix
\section{Derivative expansion: Coefficients}\label{AppA}
The explicit expressions for the coefficients in \eqref{FinalTraceT8} were obtained by the \texttt{Mathematica} tensor-algebra bundle \texttt{xAct} \cite{xAct,xTensor,xPert,Nutma2014} and read 
\begin{align}
C_0={}&Q_{2}\left[W_{(0,0,0,0)}^{(1)}\right],\label{C0}\\
C_{24}={}& 15 c_3^2 \,Q_4\left[W_{(0,0,0,0)}^{(3)}\right],\\
C_{26}={}&-15 c_3^2 \,Q_4\left[W_{(1,0,0,0)}^{(4)}\right] + 60 c_3^2 \,Q_5\left[W_{(2,0,0,0)}^{(5)}\right] - 30 c_3^2 \,Q_5\left[W_{(0,1,0,0)}^{(4)}\right],\\
C_{28}={}&15 c_3^2 \,Q_4\left[W_{(2,0,0,0)}^{(5)}\right] - 
\frac{15}{2} c_3^2 \,Q_4\left[W_{(0,1,0,0)}^{(4)}\right] - 
180 c_3^2 \,Q_5\left[W_{(3,0,0,0)}^{(6)}\right]  \nonumber\\
&+ 
180 c_3^2 \,Q_5\left[W_{(1,1,0,0)}^{(5)}\right] - 
30 c_3^2 \,Q_5\left[W_{(0,0,1,0)}^{(4)}\right] + 
720 c_3^2 \,Q_6\left[W_{(4,0,0,0)}^{(7)}\right]  \nonumber\\
&  - 
720 c_3^2 \,Q_6\left[W_{(2,1,0,0)}^{(6)}\right]+ 
120 c_3^2 \,Q_6\left[W_{(1,0,1,0)}^{(5)}\right] - 
360 c_3^2 \,Q_6\left[
W_{(2,1,0,0)}^{(6)}\right]\nonumber\\
& + 
180 c_3^2 \,Q_6\left[W_{(0,2,0,0)}^{(5)}\right] + 
120 c_3^2 \,Q_6\left[
W_{(1,0,1,0)}^{(5)}\right]- 
30 c_3^2 \,Q_6\left[W_{(0,0,0,1)}^{(4)}\right],\\
C_{36a}={}& 120 c_3 c_4 \,Q_4\left[W_{(0,0,0,0)}^{(3)}\right] + 
510 c_3^3 \,Q_5\left[W_{(0,0,0,0)}^{(4)}\right],\\
C_{36b}={}&60 c_3 c_4 \,Q_4\left[W_{(0,0,0,0)}^{(3)}\right] + 150 c_3^3 \,Q_5\left[W_{(0,0,0,0)}^{(4)}\right],\\
C_{38a}={}&93 c_3^3 \,Q_4\left[W_{(0,0,0,0)}^{(4)}\right] - 
180 c_3 c_4 \,Q_4\left[W_{(1,0,0,0)}^{(4)}\right] - 
2152 c_3^3 \,Q_5\left[W_{(1,0,0,0)}^{(5)}\right]\nonumber\\
& + 
720 c_3 c_4 \,Q_5\left[
W_{(2,0,0,0)}^{(5)}\right]- 
360 c_3 c_4 \,Q_5\left[W_{(0,1,0,0)}^{(4)}\right] + 
10760 c_3^3 \,Q_6\left[W_{(2,0,0,0)}^{(6)}\right]\nonumber\\
& - 
4304 c_3^3 \,Q_6\left[W_{(0,1,0,0)}^{(5)}\right],\\
C_{38b}={}&39 c_3^3 \,Q_4\left[W_{(0,0,0,0)}^{(4)}\right] - 
60 c_3 c_4 \,Q_4\left[W_{(1,0,0,0)}^{(4)}\right] - 
1056 c_3^3 \,Q_5\left[W_{(1,0,0,0)}^{(5)}\right] \nonumber\\
&+ 
240 c_3 c_4 \,Q_5\left[
W_{(2,0,0,0)}^{(5)}\right]- 
120 c_3 c_4 \,Q_5\left[
W_{(0,1,0,0)}^{(4)}\right] + 
5280 c_3^3 \,Q_6\left[
W_{(2,0,0,0)}^{(6)}\right]\nonumber\\
& - 
2112 c_3^3 \,Q_6\left[W_{(0,1,0,0)}^{(5)}\right],\\
C_{38c}={}& 30 c_3^3 \,Q_4\left[W_{(0,0,0,0)}^{(4)}\right] - 120 c_3 c_4 \,Q_4\left[W_{(1,0,0,0)}^{(4)}\right] - 920 c_3^3 \,Q_5\left[W_{(1,0,0,0)}^{(5)}\right] \nonumber\\
&+ 480 c_3 c_4 \,Q_5\left[W_{(2,0,0,0)}^{(5)}\right]- 240 c_3 c_4 \,Q_5\left[W_{(0,1,0,0)}^{(4)}\right] + 4600 c_3^3 \,Q_6\left[W_{(2,0,0,0)}^{(6)}\right]\nonumber\\
& - 1840 c_3^3 \,Q_6\left[W_{(0,1,0,0)}^{(5)}\right],\\
C_{48a}={}& 36 c_4^2 \,Q_4\left[W_{(0,0,0,0)}^{(3)}\right] + 594 c_3^2 c_4 \,Q_5\left[W_{(0,0,0,0)}^{(4)}\right] + 2067 c_3^4 \,Q_6\left[W_{(0,0,0,0)}^{(5)}\right],\\
C_{48b}={}&24 (3 c_4^2 + 5 c_3 c_5) \,Q_4\left[W_{(0,0,0,0)}^{(3)}\right] + 1098 c_3^2 c_4 \,Q_5\left[W_{(0,0,0,0)}^{(4)}\right] + 2694 c_3^4 \,Q_6\left[W_{(0,0,0,0)}^{(5)}\right]\\
C_{48c}={}& 12 (9 c_4^2 + 10 c_3 c_5) \,Q_4\left[W_{(0,0,0,0)}^{(3)}\right] + 2322 c_3^2 c_4 \,Q_5\left[W_{(0,0,0,0)}^{(4)}\right] + 6756 c_3^4 \,Q_6\left[W_{(0,0,0,0)}^{(5)}\right],\\
C_{48d}={}&12 (9 c_4^2 + 10 c_3 c_5) \,Q_4\left[W_{(0,0,0,0)}^{(3)}\right] + 
2322 c_3^2 c_4 \,Q_5\left[W_{(0,0,0,0)}^{(4)}\right] + 
6756 c_3^4 \,Q_6\left[W_{(0,0,0,0)}^{(5)}\right],\\
C_{48e}={}& 36 (9 c_4^2 + 10 c_3 c_5) \,Q_4\left[W_{(0,0,0,0)}^{(3)}\right] + 5076 c_3^2 c_4 \,Q_5\left[W_{(0,0,0,0)}^{(4)}\right] + 11448 c_3^4 \,Q_6\left[W_{(0,0,0,0)}^{(5)}\right],\\
C_{48f}={}& 24 (9 c_4^2 + 
10 c_3 c_5) \,Q_4\left[
W_{(0,0,0,0)}^{(3)}\right] + 
4644 c_3^2 c_4 \,Q_5\left[W_{(0,0,0,0)}^{(4)}\right]+
13512 c_3^4 \,Q_6\left[W_{(0,0,0,0)}^{(5)}\right].\label{C48f}
\end{align}
\bibliographystyle{JHEP}
\bibliography{FRGGalileon}{}
\end{document}